\documentclass[referee]{mn2e}

\usepackage{graphicx}
\usepackage{amsmath}

\title[Gamma-rays from kpc jets]
{GeV-TeV $\gamma$-rays produced by electrons in the kpc scale jet as a result of comptonization of the inner jet emission}
\author[W. Bednarek]
{W. Bednarek \\ 
Department of Astrophysics, The University of \L \'od\'z,
ul. Pomorska 149/153, 90-236 \L \'od\'z, Poland, bednar@uni.lodz.pl}
\begin{document}

\date{Accepted . Received ; in original form }

\pagerange{\pageref{firstpage}--\pageref{lastpage}} \pubyear{2015}

\maketitle

\label{firstpage}
\begin{abstract}
A few radio galaxies have been discovered to emit GeV-TeV $\gamma$-ray emission in spite of relatively large viewing angles of their jets. We consider a scenario in which such large angle $\gamma$-ray emission is
preferentially produced by isotropically distributed relativistic electrons in the kpc scale jet. These electrons comptonize soft, non-thermal radiation from the inner jet which is beamed due to its relativistic
motion. In terms of such a model, we interpret recently discovered TeV $\gamma$-ray emission
from the kpc scale jets in Cen~A. Interestingly, due to the geometry of the inverse Compton scattering process, intensity of the $\gamma$-ray emission from the counter-jet is predicted to be on a higher level than the emission from the jet. This is true provided that both, a jet and a counter-jet, are twins, i.e. the physical processes in the jet and counter-jet result in the acceleration of particles in a similar way.
However, in general, the jet and counter-jet can differ significantly. Therefore, their relative contribution to the $\gamma$-ray emission from the jets in Cen~A can become more complicated. Investigation of the distribution of the TeV $\gamma$-ray emission along jets of nearby radio galaxies with the next generation telescopes should provide constraints on the considered model and the content of the kpc-scale jets.
\end{abstract}

\section{Introduction}

Radio galaxies are commonly interpreted in terms of the blazar phenomenon  with the inner jets moving with Lorentz factors of the order of ten (e.g. Chiaberge et al 2001). However, their jets are viewed at large inclination angles. Therefore, the Doppler factors, characterising their emission proprieties, are not far from unity. The closest object of this type of active galactic nuclei (AGN), 
Centaurus~A, has the jet oriented at a large angle to the line of sight, estimated in the range $\alpha\sim (12-45)^\circ$ (M\"uller et al.~2014) and $\alpha\sim (50-80)^\circ$ (Tingay et al.~2001). On the hundred parsec distance scale, the jet in Cen A becomes subluminal with the apparent speed of $\sim 0.5c$
(Hardcastle et al. 2003). Therefore, it is expected that the jet strongly decelerates between the sub-pc and kpc distance scale.

The kpc scale jet in Cen~A is characterised by the diffusive X-ray emission (Feigelson et al.~1981) which is interpreted as being due to the synchrotron process from electrons with TeV energies (Hardcastle et al.~2006). Recently, the {\it Fermi} satellite detected the GeV $\gamma$-ray emission from the core of Cen~A (Abdo et al.~2010). The GeV spectrum is atypical, showing a hardening above $2.35\pm 0.08$~GeV (Sahakian et al.~2013, Brown et al.~2017, Sahakyan et al 2018, Abdalla et al.~2018). The TeV $\gamma$-ray emission has been also discovered from the core region of Cen~A by the HESS Collaboration (Aharonian et al.~2009). The TeV 
$\gamma$-ray spectrum, spectral index $2.52\pm 0.13_{\rm stat}\pm 0.20_{\rm sys}$ in the energy range 0.25-6~TeV (Abdalla et al.~2018), smoothly connects to the higher energy component in the GeV range. It shows no evidence of variability. Recently, it has been announced that the TeV $\gamma$-ray emission is elongated along the kpc scale jet of Cen~A (Sanchez et al.~2018).

Now, it is evident that Cen~A is a source of $\gamma$-rays on every distance scale. However, the mechanism of this emission is not clear. At first, classical blazar models have been proposed as a possible explanation of a parsec scale emission from the core, either one zone homogeneous Synchrotron self-Compton (SSC) model, e.g. Chiaberge et al.~(2001), or more advanced structured jet models (Tavecchio \& Ghiselini~2008, Giannios et al.~2009), or two zone homogeneous models (Abdalla et al.~2018). More complicated models for the large angle core emission, postulating the presence of two emission regions which exchange the radiation field, have been also proposed by Georganopoulos et al.~(2005) and Banasinski \& Bednarek~(2018). The large angle $\gamma$-ray emission in radio galaxies is also expected in the inverse Compton (IC) $e^\pm$-pair cascade models in which the geometry of the magnetic field plays an important role (e.g. Sitarek \& Bednarek~2010, or Roustazadeh \& 
B\"ottcher~2011). Finally, the large angle TeV $\gamma$-ray emission is proposed to be related to the processes occurring in the magnetosphere of the rotating super massive black hole (SMBH) (e.g. Rieger \& Mannheim~2002, Neronov \& Aharonian~2007, Rieger \& Aharonian~2008, Levinson \& Rieger~2011, Aleksi\'c et al.~2014).    

The $\gamma$-ray emission on the larger distance scale, i.e. from the kpc scale jet, likely has different origin. It has been proposed to be produced in either the comptonization process of different types of soft radiation by the TeV electrons in relativistically moving kpc scale jet (e.g. Stawarz et al.~2003, Hardcastle \& Croston~2011) or the comptonization of radiation from the stars and stellar clusters which entered the kpc scale jet (Wykes et al.~2015, Bednarek \& Banasinski~2015). In these last scenarios, electrons are accelerated on the shocks around massive stars, late type stars or other compact objects in stellar clusters. 

In this paper we investigate another comptonization scenario as a possible explanation of the TeV $\gamma$-ray emission extended along the kpc scale jets in radio galaxies. We propose that multi-TeV electrons, in the subluminal kpc jet of the radio galaxy Cen~A, inverse Compton up-scatter soft inner jet radiation into the GeV-TeV $\gamma$-ray energy range. Due to the geometry of the IC process, and a relatively low velocity of the kpc scale jet, $\gamma$ rays are preferentially produced at large angles to the jet direction. In such a case, the emission from the counter-jet should dominate over the emission from the jet. 
Similar effect has been already predicted in the inverse Compton model considerred for the X-ray emission from radio galaxies (see Brunetti, Setti \& Comastri 1997).
Proposed scenario could be tested by the future telescopes such as the Cherenkov Telescope Array (CTA, Acharya et al.~2013).

\section{Gamma-rays from kpc scale jet}

Radio galaxies, such as Cen~A, are expected to be BL Lac type active galaxies viewed at large angle to the jet direction (e.g. Chiaberge et al.~2001). Their relativistic inner jets are expected to decelerate from large Lorentz factors to the sub-relativistic velocity at the distance $>0.1$ kpc. 
In fact, observations of the kpc scale jet show many discrete features which moves with rather low velocities
(e.g. Goodger et al.~2010). The multiwavelength spectrum of Cen~A has typical for AGNs two broad bump structure. The lower bump is likely produced by electrons in the synchrotron process occurring in the inner jet. This radiation is not expected to be strongly Doppler busted due to the large inclination angle of the jet. In the case of Cen~A, this Doppler factor is assumed to be close to unity, $D_{\rm obs} = [\Gamma (1 - \beta\cos\alpha)]^{-1}\sim 1$, where $\Gamma$ is the Lorentz factor of the jet, $\beta$ is its velocity in units of the velocity of light, and $\alpha$ is the inclination angle (e.g. Chiaberge et al.~2001). However, emission along the jet axis should be relativistically boosted resulting in a much stronger 
emission along the jet than observed directly from Cen~A, due to the large Doppler factor of the inner jet. 
For the angle $\alpha = 0^\circ$, $D_{\rm j} = [\Gamma (1 - \beta)]^{-1}\approx 2\Gamma$.
In the case of Cen~A, the Doppler factor is expected to be of the order of several, for realistic values of $\alpha = 0^\circ$ and $\Gamma\sim 7$. In such a case, the soft radiation field, as observed along the large scale jet, can be dominated by the beamed radiation produced in the inner jet but not by the thermal emission from the direct vicinity of the SMBH such as an accretion disk, a broad line region (BLR) or a dusty torus (DT).

In the model considered in this paper, we assume that relativistic electrons are accelerated in the kpc scale jet. Such electrons could appear as a result of the one (or a few) known acceleration processes, either a shock acceleration, or an acceleration in turbulent plasma, or a forced reconnection of the jet magnetic field due to its collisions with obstacles in the jet (clouds, stars, clusters of stars). We do not consider specific process since they are not precisely defined in the jet environment at the present moment. These electrons are expected to be isotropic in the reference frame of the subluminal kpc scale jet. They are injected with the power law spectrum up to the maximum energy $E_{\rm max}$, $dN_{\rm e}/dE_{\rm e} = AE_{\rm e}^{-\delta}$, where $A$ is the normalization constant and $\delta$ is the spectral index. Since these electrons are at the kpc distance scale from the inner jet, they see the inner jet soft radiation almost mono-directional, i.e. from the direction of the SMBH. Note that the scenario proposed above has many common features to that one already considered for the radiation from the two decelerating blobs moving along the inner jet (see 
Banasi\'nski \& Bednarek~2018). However, in the present case electrons are located at large distance from the jet base (at a kpc scale). At such distances, the jet becomes already non-relativistic. Therefore, in the present model it is expected that, due to the geometrical effects, the TeV $\gamma$-ray emission will be preferentially produced at very large angles to the jet axis. The schematic picture of the model is presented in Fig.~1. 

\begin{figure}
\vskip 4.8truecm
\includegraphics{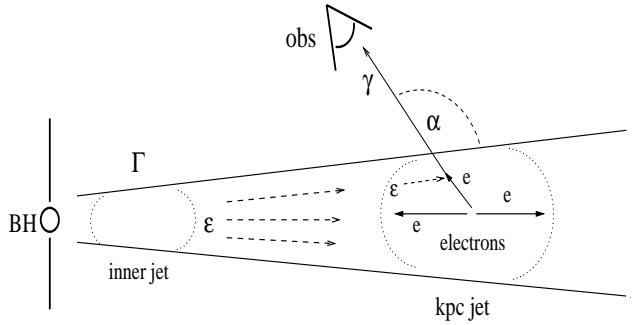}
\caption{Schematic presentation of the $\gamma$-ray production by relativistic electrons (e) in the kpc scale jet which comptonize the non-thermal, soft radiation produced in the inner, relativistic jet. The inner jet moves with the Lorentz factor $\Gamma$. Electrons have isotropic distribution in the kpc-scale jet which moves semi-relativistic along the jet. Non-thermal, soft radiation is produced in the inner, relativistic jet. 
$\gamma$-rays are preferentially produced towards the observer located at the large angle, $\alpha$, in respect to the direction defined by the jet axis.}
\label{fig1}
\end{figure}

In order to calculate the $\gamma$-ray spectra produced by electrons,  we have to determine the mono-directional soft radiation field, produced in the inner jet, as observed in the outer jet.  
We derive such radiation field in the case of the jet in Cen~A from its observed soft radiation flux at the observer, $F(\varepsilon)$ [ph. cm$^{-2}$s$^{-1}$MeV$^{-1}$]. We have used the data obtained by Marconi et al. (2000), which base on the WFPC2 and NICMOS observations, and the data obtained by the IRAS, SCUBA and ISOCAM from Mirabel et al. (1999). These data are collected in Fig.~6 in Marconi et al.~(2000), see also Fig.~1 in Chiaberge et al.~(2000). We consider this spectrum as a lower limit on the non-thermal synchrotron emission from the inner region in the jet of Cen~A. For our purposes, the photon flux of the low energy bump in the Cen~A spectrum is approximated by the combination of three power law functions,
\begin{eqnarray}
F(\varepsilon)~~\left[{{\rm ph.}\over{{\rm MeV}{\rm cm}^2{\rm s}}}\right] = 
\begin{cases}
A_1\varepsilon^{-\beta_1} & {\rm if}~~\varepsilon < \varepsilon_1;\\
A_2\varepsilon^{-\beta_2} & {\rm if}~~ \varepsilon_1 < \varepsilon < \varepsilon_2;\\
A_3\varepsilon^{-\beta_3} & {\rm if}~~\varepsilon > \varepsilon_2.
\end{cases}
\label{eq1}
\end{eqnarray}
\noindent
where the normalization factors are equal to $A_1 = 8.82\times 10^{9}$~MeV$^{-1}$~cm$^{-2}$~s$^{-1}$, $A_2 = 1.48\times 10^{-4}$~MeV$^{-1}$~cm$^{-2}$~s$^{-1}$, $A_3 = 1.95\times 10^{-52}$~MeV$^{-1}$~cm$^{-2}$~s$^{-1}$, the energy breaks are at 
$\varepsilon_1 = 1.3\times 10^{-9}$ MeV and $\varepsilon_2 = 4.13\times 10^{-7}$ MeV, and the spectral indexes are $\beta_1 = 0.45$, $\beta_2 = 2.0$, and $\beta_3 = 9.5$, and $\varepsilon$ is the photon energy as observed by the distant observer. 

The differential density of soft photons from the inner jet at the location of the outer jet at the distance, $z$, along the jet is,
\begin{eqnarray}
n(\varepsilon') = F(\varepsilon)d_{\rm L}^2D_{\rm j}^2/z^2c,
\label{eq2}
\end{eqnarray}
\noindent
where $d_{\rm L} = 3.8$ Mpc is the luminosity distance to Cen~A (Harris et al.~2010), $D_{\rm j}$ is the Doppler factor of the inner jet, the busted photon energy in the kpc jet is $\varepsilon' = \varepsilon D_{\rm j}$,  and $c$ is the velocity of light. Note that we assumed that the observed radiation from Cen~A is not Doppler busted, i.e. $D_{\rm obs} = 1$.

We calculate the $\gamma$-ray spectra produced by isotropic electrons with the equilibrium spectrum, 
$dN_{\rm e}/dE_{\rm e}$, which scatter mono-directional soft radiation with the spectrum, $n(\varepsilon)$,
by integration of the formula Eq.~3 in Banasi\'nski \& Bednarek~(2018).
As an example, we calculate the $\gamma$-ray spectra produced by electrons as a function of the observation angle, $\alpha$. Electrons are assumed to have the power law spectrum with the spectral index $\delta = 3$ (Fig.~2a) and $\delta = 2$ (Fig. 2b) up to 30 TeV. Electrons are at the distance of $z = 1$~kpc from the base of the jet. The inner jet moves with the Doppler factor $D_{\rm j} = 15$. Note the interesting dependence of the $\gamma$-ray emission on the observation angle $\alpha$ (Fig.~2a,b). It becomes the strongest in directions towards the inner jet, i.e. for $\alpha = 180^\circ$. On the other hand, $\gamma$-ray emission along the jet axis drops to zero in such comptonization scenario. 
We also investigate the dependence of the $\gamma$-ray spectra on the maximum energy of relativistic electrons within the kpc scale jet (Fig.~2c). The Klein-Nishina effect on the $\gamma$-ray spectrum is clearly seen in this figure for the electrons which spectra show the cut-off at large energies. 
Finally, we consider the dependence of the $\gamma$-ray spectrum on the spectral index of electrons in the kpc scale jet. They peak at low energies (for the steep electron spectra) or at the high energies (for the flat spectra). In summary, the $\gamma$-ray spectra produced in this IC model show unique features. They should allow to distinguish this model from other IC models. The most interesting feature is the dominant $\gamma$-ray emission from the counter-jet (observed at the angle $\alpha > 90^\circ$) in respect to the $\gamma$-ray emission from the jet (observed at the angle $\alpha < 90^\circ$). In the next chapter, we apply such scenario to recent $\gamma$-ray observations of the nearby radio galaxy Cen~A. 

\begin{figure*}
\vskip 4.3truecm
\includegraphics{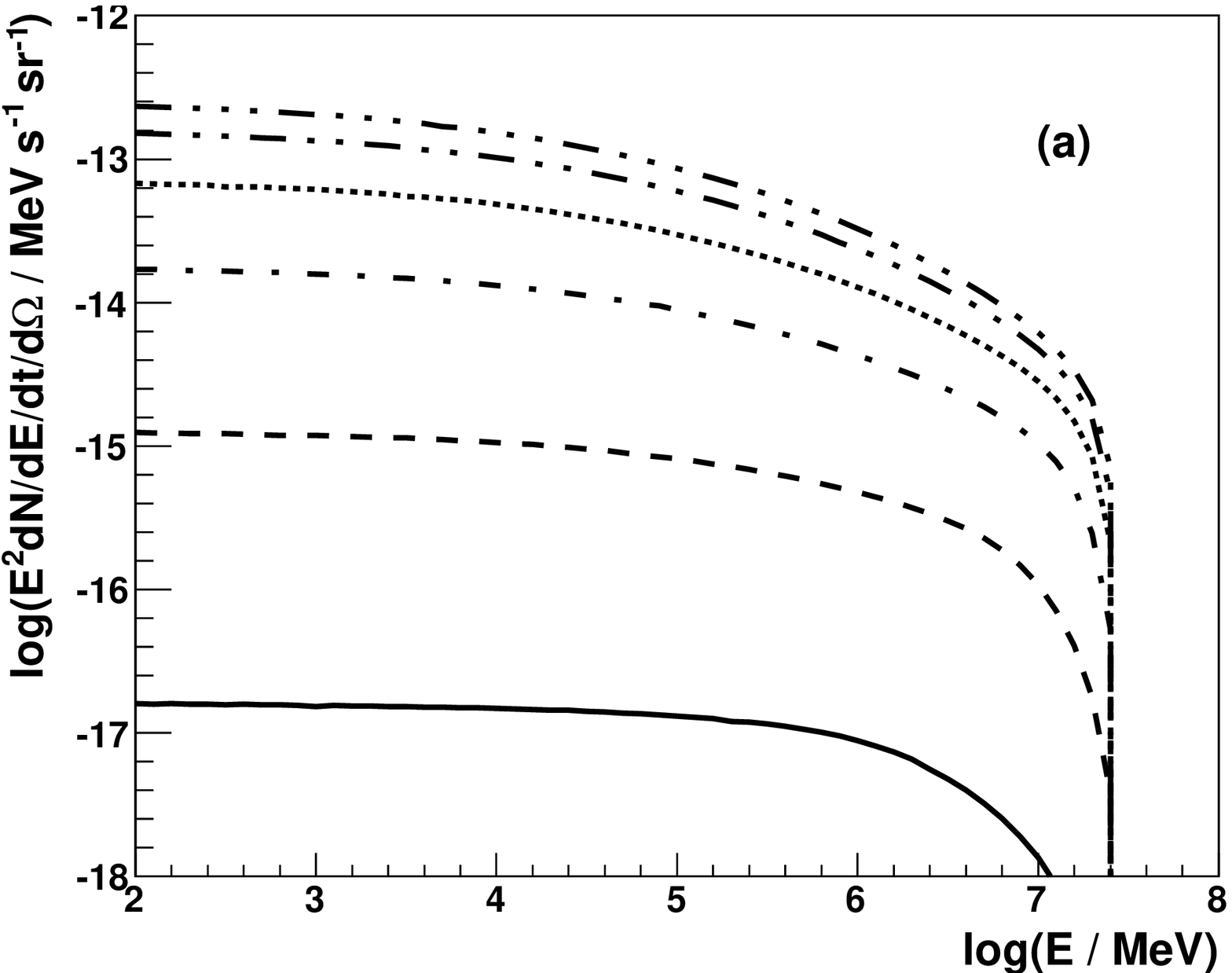}
\includegraphics{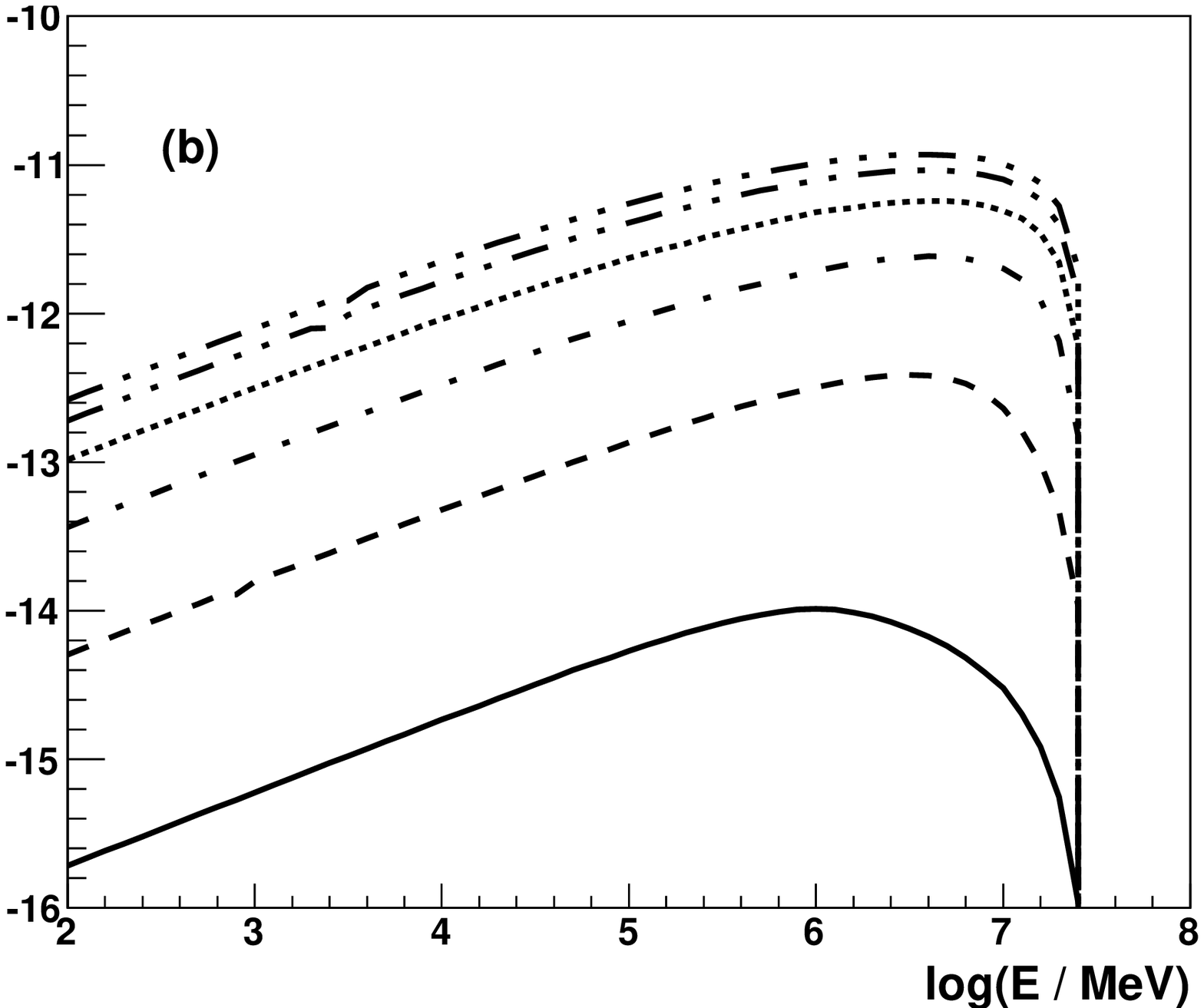}
\includegraphics{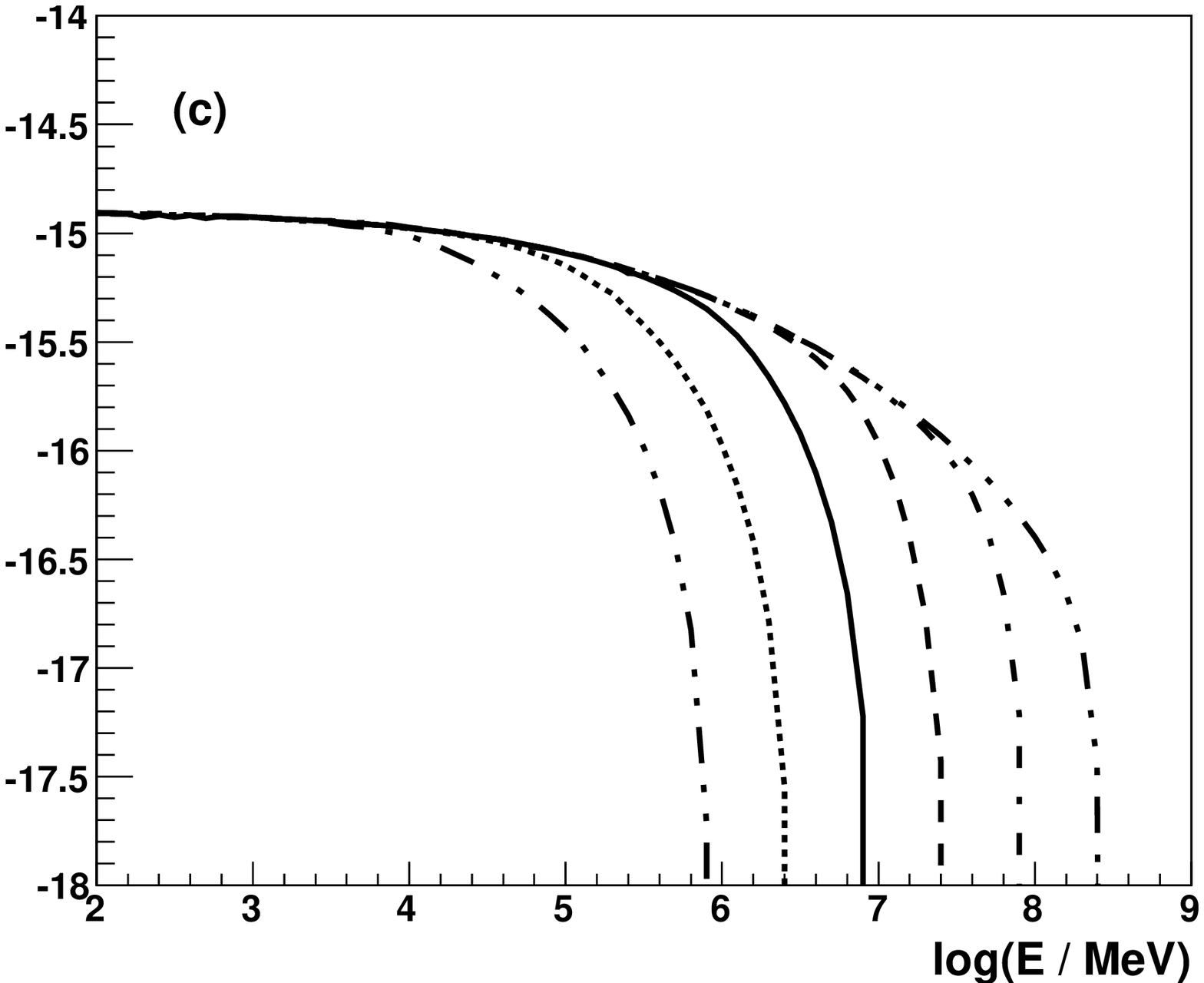}
\includegraphics{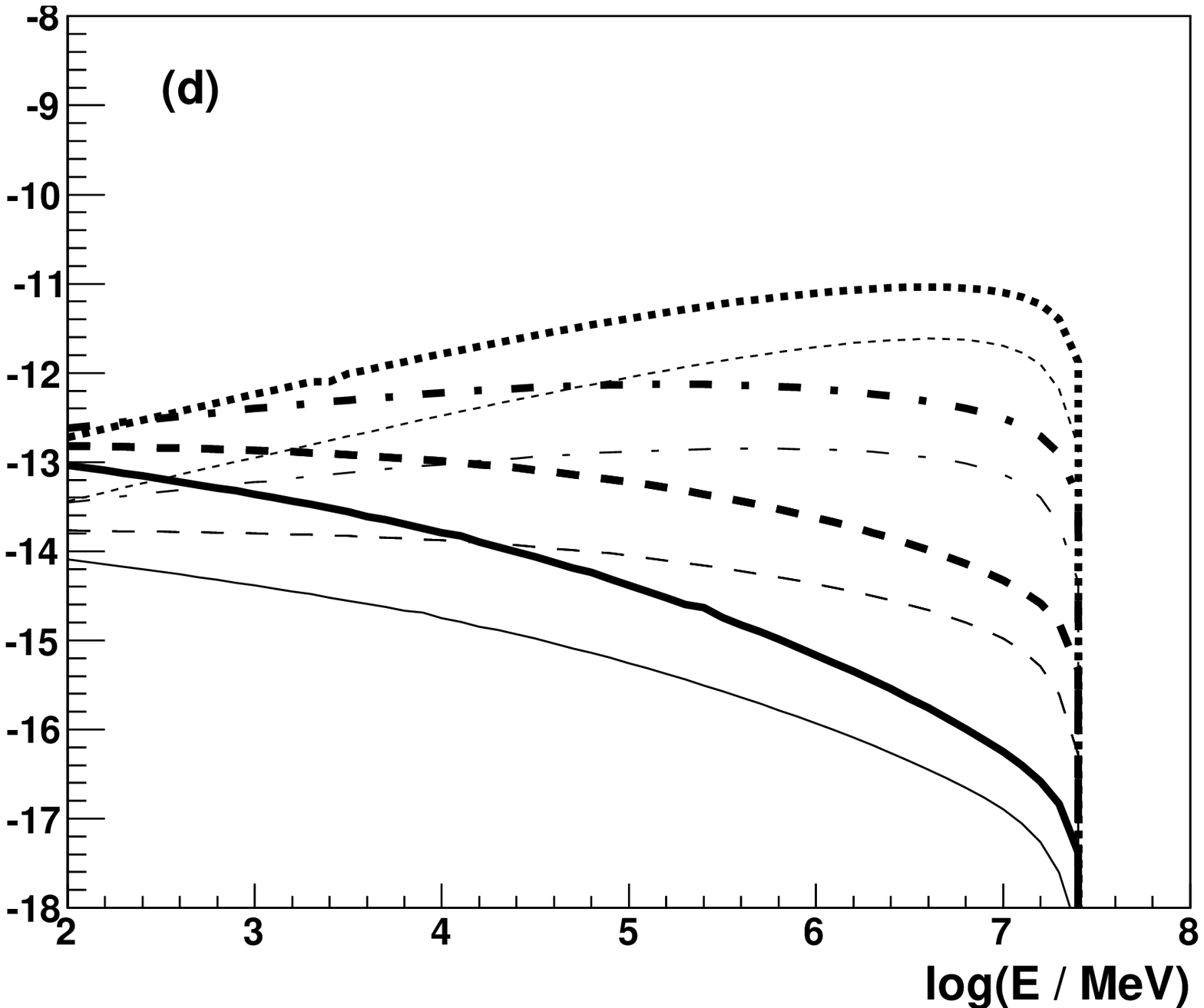}
\caption{Spectral Energy Distribution (SED) of the $\gamma$-rays produced by relativistic electrons, in the kpc scale jet, which comptonize soft, non-thermal radiation produced in the inner jet. The inner jet moves with the Doppler factor $D_{\rm j} = 15$. Electrons are injected isotropically at the distance of 1 kpc from the jet base. They are characterised by the power law spectrum with the spectral index $\delta$ between $E_{\rm e}^{\rm min} = 1$~GeV and $E_{\rm e}^{\rm max} = 30$~TeV. The total energy in this spectrum is normalized to 1 MeV. Dependence of the SED on the observation angle 
$\alpha = 10^\circ$ (solid), 30$^\circ$ (dashed), $60^\circ$ (dot-dashed), 90$^\circ$ (dotted), 120$^\circ$ (dot-dot-dashed), 150$^\circ$ (dot-dot-dot-dashed) is shown for $\delta = 3$ (figure a) and $\delta = 2$ (figure b). Dependence of SED on the maximum energy in electron spectrum, $E_{\rm e}^{\rm max} = 1$~TeV (dot-dot-dashed), 3 (dotted), 10 (solid), 30 (dashed), 100 (dot-dashed), and 300 TeV (dot-dot-dot-dashed) is shown in (c) for $\delta = 3$, and dependence on the electron spectral index  
$\delta = 2$ (dotted), 2.5 (dot-dashed), 3 (dashed), and 3.5 (solid) for the observation angles equal to 
$\alpha = 120^\circ$ (thick curves) and $60^\circ$ (thin) is shown in (d). The spectrum of soft radiation from the inner jet is shown in Fig.~3 by the solid curve.}
\label{fig2}
\end{figure*}
\section{TeV $\gamma$-rays from kpc jet in Cen A}

{\it Chandra} observations show diffusive non-thermal X-ray emission extending up to 7 keV which dominates over X-ray emission from the knots in the jet of Cen~A (Hardcastle et al.~2003, Hardcastle et al.~2006). If this emission is interpreted as due to the synchrotron process, then the acceleration efficiency of electrons in the kpc scale jet, $\xi$, can be constrained provided that the acceleration process of electrons is saturated by their synchrotron energy losses. The following relation allows to estimate the energy of synchrotron photons produced by electrons accelerated in the radiation site,
\begin{eqnarray}
\varepsilon_{\rm syn}\approx 1.5\times 10^5\xi~~~{\rm keV}.
\label{eq3}
\end{eqnarray}
\noindent
By reversing the above formula, we estimate $\xi$ on $\sim 5\times 10^{-5}$.
On the other hand, we can also roughly estimate the strength of the magnetic field at the kpc jet by extrapolating it from the vicinity of the central engine. We assume that the Poynting flux through the jet, 
at its base, is of the order of the jet power. In the case of Cen~A, this power has been estimated on 
$L_{\rm j}\sim$10$^{43}$ erg~s$^{-1}$ (Wykes et al.~2013). Then,
\begin{eqnarray}
L_{\rm j} = L_{\rm P} = \pi R_{\rm in}^2c(B_{\rm in}^2/8\pi)\gamma_{\rm j}^2, 
\label{eq4}
\end{eqnarray}
\noindent
where $R_{\rm in} = 3R_{\rm Sch}$ is the inner radius of the jet and $R_{\rm Sch}$ is the Schwartzschild radius of the black hole with the mass $(5.5\pm 3)\times 10^7$~M$_\odot$, as estimated in the case of Cen~A (Cappellari et al.~2009), $\gamma_{\rm j}$ is the jet Lorentz factor assumed to be of the order of 7. From the above relation, we estimate the magnetic field strength at the base of the jet, $B_{\rm in}\sim 200$ Gs. In a simple conical jet model, the dominant toroidal component of the magnetic field should drop according to, $B_{\rm kpc}\sim B_{\rm in}R_{\rm in}/R_{\rm kpc}\approx 3.7\times 10^{-6}$~Gs, where $R_{\rm kpc}$ is the distance along the jet equal to 1 kpc. This estimate should be considered as the lower limit. For example, in the case of the parabolic jet, the magnetic field in the kpc scale jet should be clearly stronger.

Above determined physical parameters in the kpc jet of Cen A allow us to estimate the maximum energies to which electrons can be accelerated in the presence of dominant synchrotron energy losses,
\begin{eqnarray}
E_{\rm e}^{\rm max}\approx 55 (\xi/B_{\rm kpc})^{1/2}\approx 200~~~{\rm TeV}.
\label{eq5}
\end{eqnarray}
\noindent
We conclude that in order to explain observations of the non-thermal X-ray emission from the kpc scale jet in 
Cen~A, electrons have to be accelerated to energies of the order of a few hundred TeV.

\begin{figure}
\vskip 4.8truecm
\includegraphics{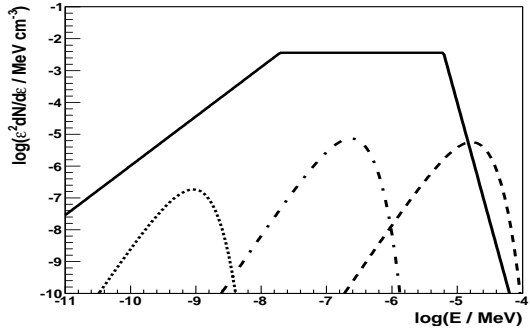}
\caption{Differential density of different types of soft radiation (multiplied by energy squared) at the distance of 1 kpc along the jet in Cen~A produced by the accretion disk (dashed), the molecular torus (dot-dashed), the inner jet (solid), and the photon density of the MBR (dotted). It is assumed that the accretion disk has the inner radius $5\times 10^{13}$ cm (corresponding to the 3 Schwarzschild radii of the black hole with the mass $5\times 10^7$~M$_\odot$) and the luminosity $10^{43}$ erg~s$^{-1}$ (dashed curve). 
Its radiation is approximated by the black body spectrum. The torus radiation, the temperature 700 K and the radius of 3 pc, is diluted by a factor of $10^{-3}$. It is assumed that the Doppler factor along the jet axis ($\alpha = 0^\circ$) is $D_{\rm j} = 15$.}
\label{fig3}
\end{figure}

Electrons with such energies are expected to produce TeV $\gamma$-rays in the scenario discussed in Sect.~2.
We calculate the GeV-TeV $\gamma$-ray spectra for the parameters of the radio galaxy Cen~A, assuming that electrons are accelerated with the power law spectrum (spectral index $\delta$) up to $E_{\rm max} = 50$ TeV. 
We have checked that defined in Sect.~2, non-thermal radiation field from the inner jet of Cen~A dominates at low energies over unbeamed soft radiation emitted from the central region of Cen~A. Such unbeamed soft radiation can be produced, either by the accretion disk around SMBH in Cen~A or the molecular torus (characteristic temperature 700 K, typical radius of 3 pc). The radiation of the torus is produced by dust. 
It is assumed to be a grey black body diluted by a factor of $10^{-3}$ corresponding to 7.8 magnitude (Marconi et al.~2000). The radiation from the accretion disk has the luminosity comparable to the jet power, i.e. $L_{\rm D} = 10^{43}$ erg~s$^{-1}$. For the known mass of the SMBH in Cen A ($5\times 10^7$~M$_\odot$), we estimate the inner radius of the accretion disk and the characteristic disk temperature at the inner radius. The radiation from the torus and the accretion disks are taken to be diluted at the 1 kpc distance scale by a factor determined by the characteristic dimensions of their production regions, i.e the radius of the torus and the inner radius of the accretion disk. Note that in the case of the optically thin disk models
(Narayan \& Yi~1995), 
the characteristic energies of the disk photons are expected to be larger. In such a case the scattering of such radiation will be less efficient due to the Klein-Nishina effects. 
The non-thermal radiation field, defined by Eq~2, also dominates over the Microwave Background Radiation (MBR) field at the kpc distance scale along the jet. We compare the applied non-thermal radiation field produced in the inner jet (but calculated at the kpc distance scale along the jet) with other radiation fields in Fig.~3. Clearly, the non-thermal inner jet radiation dominates over other radiation fields in the kpc jet below 10 eV. However, at larger energies the non-thermal radiation can be over-come by the emission from the luminous accretion disk. These most energetic optical photons (with energies $>$10~eV) do not contribute essentially to the TeV $\gamma$-ray spectrum since they mainly interact in the Klein-Nishina regime.

\begin{figure*}
\vskip 5.2truecm
\includegraphics{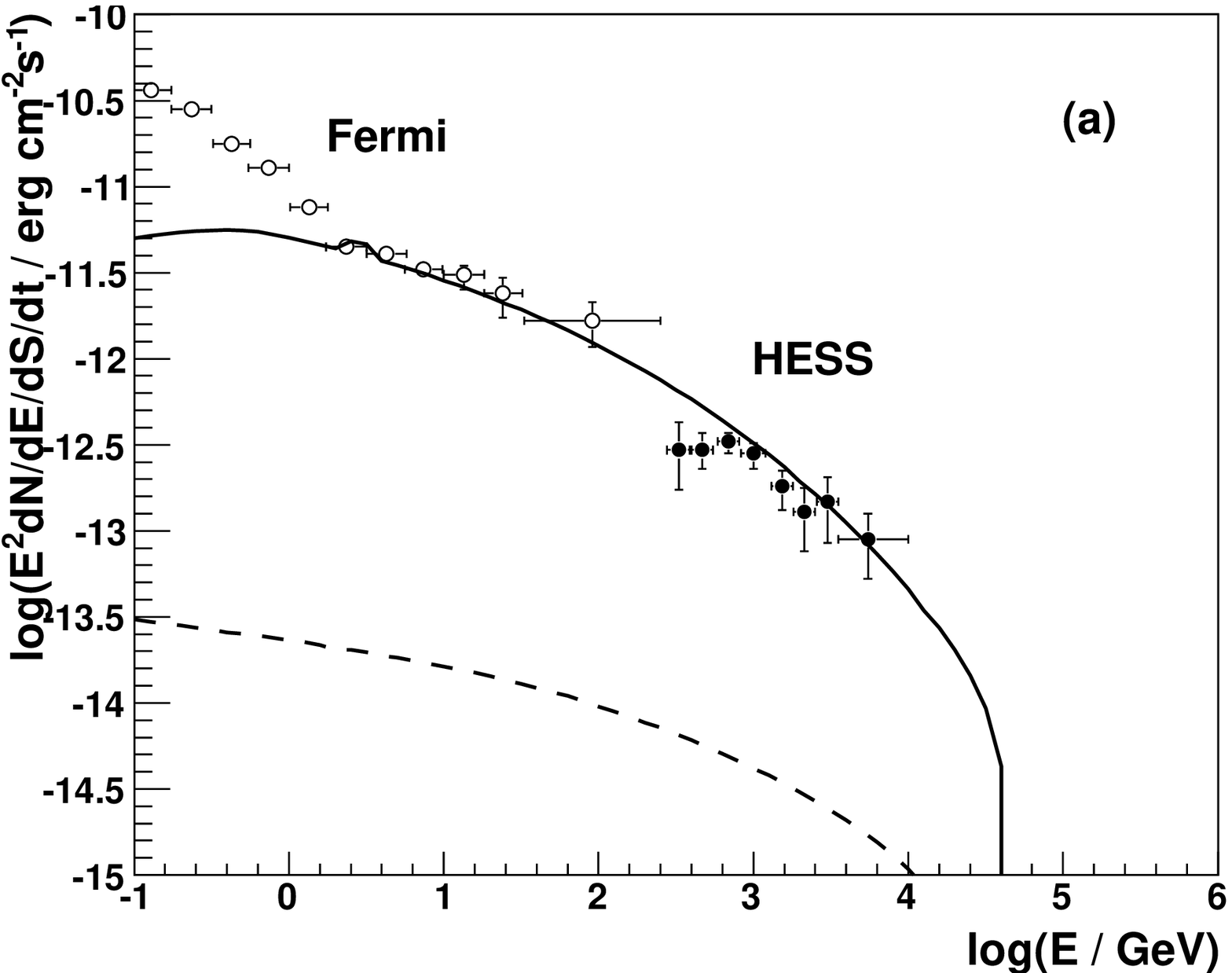}
\includegraphics{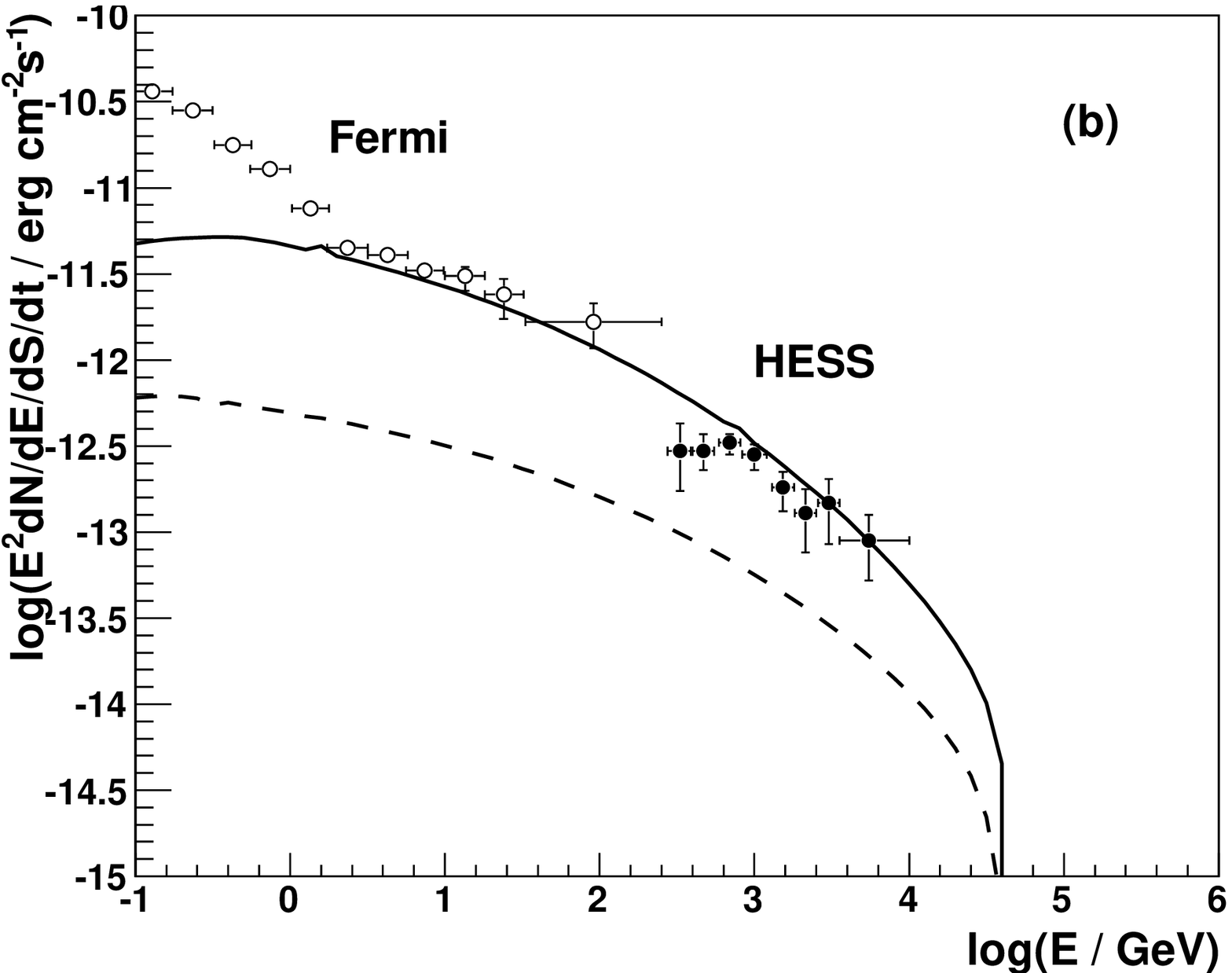}
\includegraphics{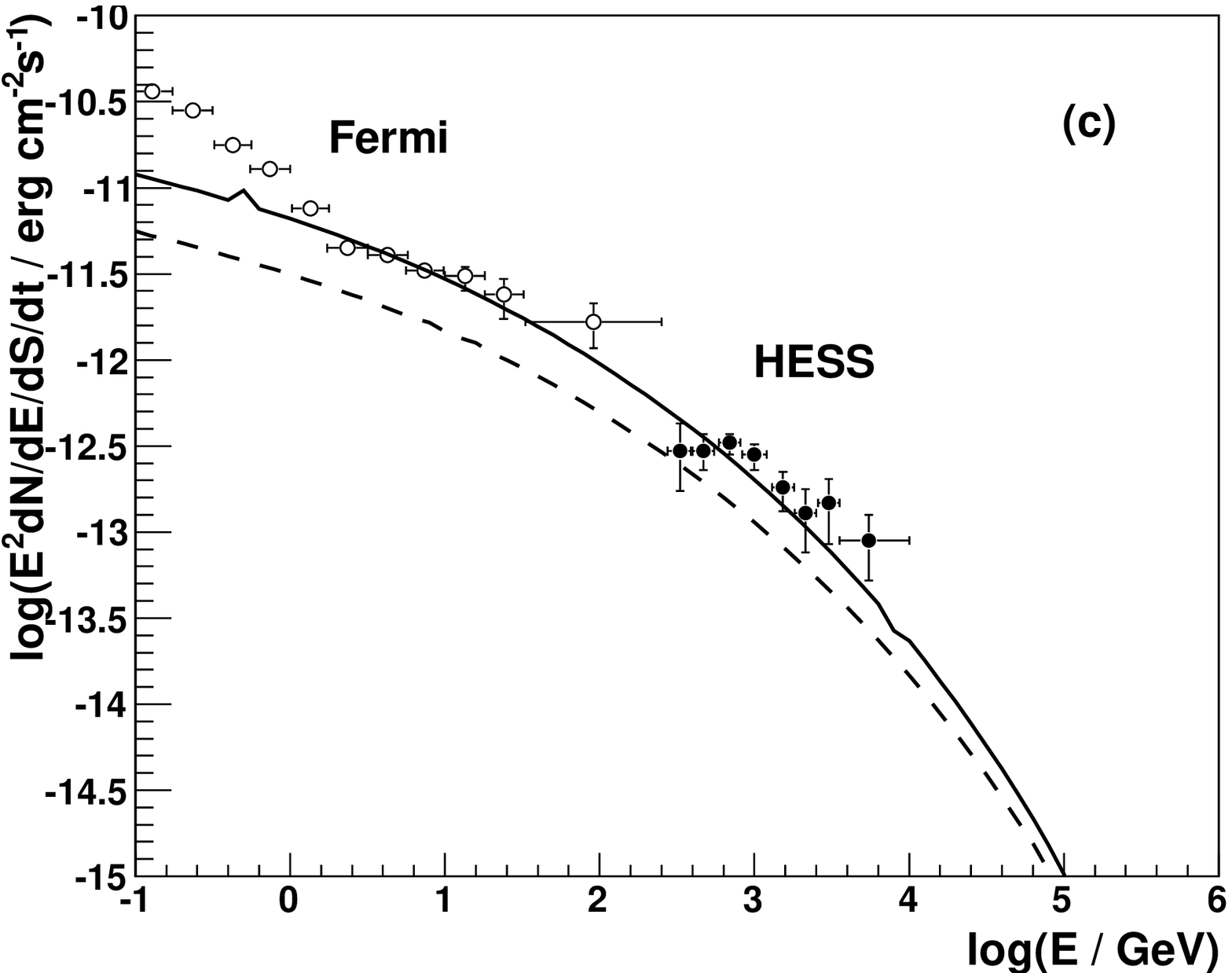}
\caption{SED of the Cen~A as measured in $\gamma$-ray energies by the Fermi-LAT Telescope (open circles) and the HESS Cherenkov telescope system (filled circles) reported by Abdalla et al. (2018). It is confronted with the predictions of the $\gamma$-ray emission produced by relativistic electrons in the kpc scale jet. These electrons comptonize the non-thermal soft radiation produced in the inner part of the jet in Cen~A (see dashed curve) and the counter-jet (see solid curve). The jet is inclined towards the observer at the angle $\alpha = 30^\circ$ (figure a), $\alpha = 60^\circ$ (b), and $\alpha = 80^\circ$ (c). 
The counter-jets are inclined at the angle $\alpha = 150^\circ$, $120^\circ$, and $100^\circ$, respectively. It is assumed that the physical proprieties in the jet and counter-jet are these same. Electrons are injected into the jets with the power law spectrum and the spectral index $\delta = 3.2$ between 3 GeV and  50 TeV (figure a and b), and $\delta = 3.4$ between 3 GeV and 200 TeV (c). The spectrum of the soft non-thermal radiation in the kpc scale jets, that is produced in the inner jets, is shown in Fig.~3. It is assumed that the inner jets move with the velocity corresponding to the Doppler factor along the jet axis equal to $D_{\rm j} = 15$.}
\label{fig4}
\end{figure*}

For such background radiation field, we calculate the $\gamma$-ray spectra produced by the isotropic electrons at the distance of 1 kpc along the jet in Cen~A. Good description of the Cen~A TeV $\gamma$-ray spectrum (and also the harder component of the GeV spectrum) is obtained for the broad range of the inclination angles of the jet in respect to the line of sight (see Fig.~4). As an example, we show three example modellings of the Cen~A spectrum, for the inclination angle of the jet equal to $\alpha = 30^\circ$ (the counter-jet seen at 
$\alpha = 150^\circ$, see figure (a)), $\alpha = 60^\circ$ (counter-jet at $\alpha = 120^\circ$ (b)), and $\alpha = 80^\circ$ (counter-jet at $\alpha = 100^\circ$ (c)). In fact, the observations are well described by the emission which comes mainly from the counter-jet. In the case of the model with the angle $\alpha = 30^\circ$, emission from the jet in the direction towards the observer is negligible in respect to the emission from the counter-jet. This intriguing emission feature of the considered model, i.e. more effective comptonization of soft radiation by electrons in the counter-jet than in the jet towards the observer, is due to the specific geometry of the IC process. Note, that such a feature is only expected when the jet and counter-jet have similar physical proprieties. This does not need to be always true since both jets can in principle differ significantly due to the accretion instabilities onto the SMBH or the environmental differences of specific jets, such as the content of compact objects within the jet/counter-jet (clouds, massive stars, or stellar clusters). These compact objects might produce turbulence in the jet providing good conditions for the acceleration process of electrons with different efficiencies in specific jets.
On the other hand, for the inclination angles closer to $90^\circ$, the contribution to the observed $\gamma$-ray spectrum from the jet and counter-jet becomes comparable. Therefore, depending on the inclination angle of the jet, either emission from both, the jet and the counter-jet is observed, or emission from only the counter-jet is observed. This interesting high energy emission feature can be investigated in detail already with the present Cherenkov telescopes. Very recently, the HESS Collaboration reported that the TeV $\gamma$-ray emission in Cen~A is extended along the kpc scale jet (Sanchez et al.~2018). 

In order to get correct description of the GeV-TeV $\gamma$-ray observations in terms of our scenario, 
the total energy in relativistic electrons has to be of the order of $E_{\rm e}\sim 10^{54}$~erg. This total energy simply scales with the Doppler factor of the inner jet, $D_{\rm j}$, the Doppler factor of the jet as seen towards the observer at the angle $\alpha$, and the distance of the emission region from the base of the jet, $z$, according to the following prescription 
$\propto (zD_{\rm obs}/D_{\rm j})^2$. On the other hand, the total energy in the kpc scale jet can be estimated on $E_{\rm j}^{\rm tot}\sim L_{\rm j} z/(\beta_{\rm j}c)\sim 10^{55}$ erg for $L_{\rm j} = 10^{43}$ erg s$^{-1}$, z = 1 kpc, and the velocity of the subluminal kpc scale jet of the order 10$\%$ of the velocity of light, $\beta_{\rm j}\sim 0.1$. Therefore, the efficiency of electron acceleration in the jet should be of the order of $10\%$.

\section{Conclusion}

We show that comptonization of the soft radiation, produced  in the inner relativistic jet, by relativistic electrons in the subluminal kpc-scale jet can be responsible for the large angle $\gamma$-ray emission in radio galaxies. In such scenario, the inner jet radiation is almost mono-directional as seen by the isotropically distributed  electrons at the kpc jet. The intriguing feature of such a model is the dominant contribution to the over-all $\gamma$-ray spectrum from the counter-jet over the contribution from the jet in the case of jets aligned at intermediate angles. This feature is due to the specific geometry of the soft radiation field, which is assumed to be mono-directional along the jet axis due to the relativistic beaming of the inner jet. On the other hand, relativistic electrons are distributed isotropically in the subluminal kpc-scale jet. With such geometry, $\gamma$-rays are produced in the IC process predominantly at large angles to the jet direction.

As an example, we propose such scenario as a possible explanation of the GeV-TeV $\gamma$-ray emission from Cen~A. Interestingly, the HESS Collaboration (Sanchez et al.~2018) recently reported that the TeV 
$\gamma$-ray emission is extended along the kpc-scale jet in Cen~A. Such intriguing emission feature can be naturally explained in terms of the considered here model. The basic prediction of the model, stronger TeV 
$\gamma$-ray emission from the counter-jet than from the jet, can be tested by 
the future CTA observations of the nearby radio galaxies. Note however, that the conditions in/around the jet and counter-jet can differ significantly. In such a case, the predicted emission in the TeV $\gamma$-ray range from the jet/counter-jet may show different dependences than predicted by the simple twin-jets scenario. 
In fact, significant differences between both jets in Cen~A are observed on a larger distance scale in the GeV $\gamma$-ray energy range produced in the radio lobes (Abdo et al.~2010). 
Then, the measurements of the TeV $\gamma$-ray emission, separately from the jet and the counter-jet, can provide interesting information on the launching mechanism of the jets and/or on the role of the surrounding medium on the acceleration process of particles in AGN jets.

\section*{Acknowledgements}
I would like to thank the Referee for useful comments.
This work is supported by the grant through the Polish Narodowe Centrum Nauki No. 2014/15/B/ST9/04043.


\label{lastpage}
\end{document}